\documentclass[review]{elsarticle}
\usepackage[numbers]{natbib}

\RequirePackage{xspace}
\RequirePackage{amsmath}
\RequirePackage{amsfonts}
\RequirePackage{amssymb}

\def\ifundefined#1{\expandafter\ifx\csname#1\endcsname\relax}

\def\la{\mathrel{\hbox{\rlap{\hbox{\lower4pt\hbox{$\sim$}}}\hbox{$<$}}}}
\def\ga{\mathrel{\hbox{\rlap{\hbox{\lower4pt\hbox{$\sim$}}}\hbox{$>$}}}}

\newcommand{\be}{\begin{equation}}
\newcommand{\ee}{\end{equation}}

\newcommand{\bea}{\begin{eqnarray}}
\newcommand{\eea}{\end{eqnarray}}

\ifundefined{ensuremath}\def\ensuremath#1{\relax\ifmmode{#1}}
\else${#1}$\fi\else\relax\fi
\ifundefined{nuc}\def\nuc#1#2{\relax\ifmmode{}^{#1}{\protect\mathrm{#2}}
\else${}^{#1}$#2\fi}\else\relax\fi

\newcommand{\etal}{et al.\xspace}
\newcommand{\gcm}{g~cm$^{-3}$\xspace}
\newcommand{\kmps}{\ensuremath{\mathrm{km}~\mathrm{s}^{-1}}\xspace}

\newcommand{\msol}{\ensuremath{{\mathrm{M}_\odot}}\xspace}

\newcommand{\nni}{\ensuremath{\nuc{56}{Ni}}\xspace}

\newcommand{\Mch}{\ensuremath{\mathrm{M}_{\mathrm{Ch}}}\xspace}

\usepackage{aas_macros}
\graphicspath{{./}{./figs/}{../lbl07/}{../figs/}{../../sn01ay/figures/}{../pitt12/}{../../sn2011fe/figs/}{../../prop/nsf/nsf_collab13/figures/}{../../sn12Z/}}

\journal{Nuclear Physics A}

   \biboptions{square,numbers,sort&compress}

\begin{document}

\begin{frontmatter}

\title{SNe~Ia: Can Chandrasekhar Mass Explosions Reproduce
  the Observed Zoo?}

\author{E.~Baron}
\address{Homer L.~Dodge Dept.~of Physics \& Astronomy, University of
  Oklahoma, 400 W.~Brooks, Rm 100, Norman, OK 73072-2061, USA}
\address{Hamburger Sternwarte, Gojenbergsweg 112, 21029 Hamburg, Germany
}

\begin{abstract}
The question of the nature of the progenitor of Type Ia supernovae
(SNe~Ia) is important both for our detailed understanding of stellar
evolution and for their use as cosmological probes of the dark
energy. Much of the basic features of SNe~Ia can be understood
directly from the nuclear physics, a fact which Gerry would have
appreciated. We present an overview of the current observational and
theoretical situation and show that it not incompatible with
most SNe~Ia being the results of thermonuclear explosions near the
Chandrasekhar mass.
\end{abstract}


\end{frontmatter}

\begin{figure}
\centering
  \includegraphics[scale=0.5]{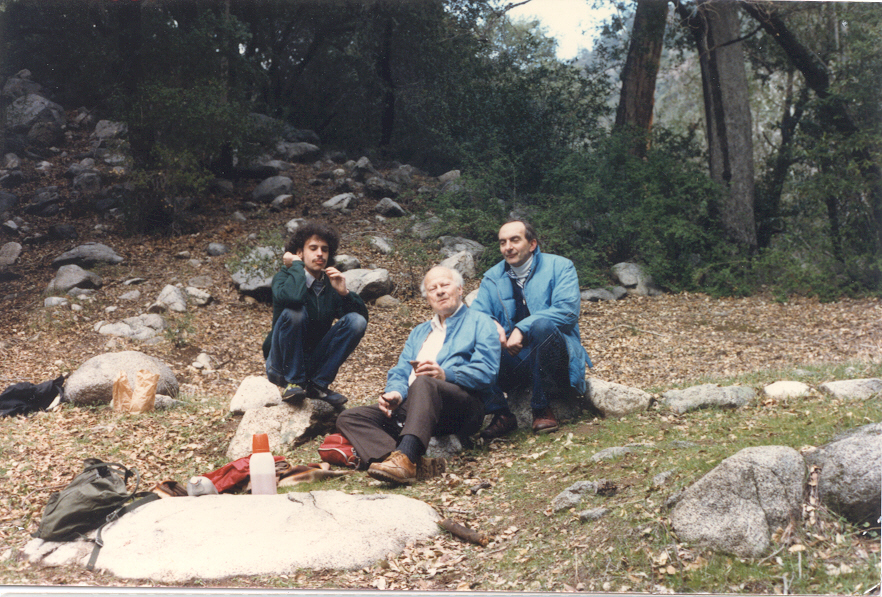}
\caption{A Saturday hike in the San Gabriel Mountains above Pasadena
  in 1985. Left to right are: E.~Baron, Hans Bethe, and Gerry Brown. Photo
  credit: Jerry Cooperstein. \label{fig:BBB85}
}
\end{figure}

\section{Introduction}

I have been fascinated listening to all of the talks and the
remembrances of Gerry. It is especially interesting to see the wide
scientific range of all of the Nuclear Theory Group alumni.  My relationship with Gerry was complex. Unlike
most of Gerry's students whom he actively recruited, when I asked Gerry
if I could work with him, he was noncommittal. I had done just okay in
Tom Kuo's Quantum Mechanics course and poorly on the QM section of the
qualifier and 
I'm sure he used that as a filter to decide which students to take
on. Nevertheless, Gerry did give me a chance and he helped me through
many of the bewildering aspects of graduate school. Actually, once I joined
the Nuclear Theory Group as a graduate student I worked much of the time
with Jerry Cooperstein (Coop) and our work preceded quite
well. When a promised fellowship fell through, I had to scramble to
find my first postdoc. I returned to the Nuclear Theory Group for my
second postdoc. In a weak economy, finding a permanent position was difficult
 and Gerry worked hard on my behalf. In the end, Gerry always came through
for me. Gerry's strong sense of fair play that many of us have
remarked on, definitely 
worked on my behalf. 

I also want to take a minute to discuss our collaborators on the
supernova problem during my time at Stony Brook. First I want to
mention the role that Coop played in both my and Gerry's scientific
work on the core collapse problem. Gerry trusted Coop implicitly. If
Coop said it then Gerry took it seriously. And while Gerry was my
thesis adviser, my day-to-day interactions were with Coop. It is
indeed a shame that Coop couldn't make it to this meeting.

The other person to mention is, of course, Hans Bethe. Gerry's
collaboration with Hans was really important to him. He was proud that
Hans was his collaborator.  Gerry, Hans, Coop, and I certainly
enjoyed the January 
``breaks'' at Caltech, Santa Barbara, and Santa Cruz (Fig~\ref{fig:BBB85}). 

\section{SNe Ia Basics}

Type Ia supernovae as observational phenomena are exceedingly regular,
particularly when compared 
with the much more diverse class of core collapse supernovae. In
astronomer's units the maximum brightness of SNe~Ia in the $B$ band,
$M_B$ is -19.25 with a 1-$\sigma$ dispersion of 0.50~mag. For ordinary
SNe IIP $M_B = -16.75$ with a 1-$\sigma$ dispersion of 0.98~mag
\citep{dean14}. This regularity led quickly to the understanding that
the progenitors of SNe~Ia were likely the thermonuclear explosion of
near Chandrasekhar mass white dwarfs \citep{ColgateMckee69}. 

In fact
the energy source of the visible display of SNe~Ia and that of
traditional core-collapse supernovae are very different. In core
collapse the underlying energy source is gravitational potential
energy, which is released during the collapse of the iron core of a
massive star to become a proto-neutron star. From an astronomical
viewpoint, the core 
collapse display, that is the observed light curve and spectra in the
UV+Optical+IR (UVOIR) is for the most part powered by energy deposited
by the shock and stored in the thermal and
ionization energy of hydrogen and other elements. 

\begin{figure}
\centering
\includegraphics[scale=0.5]{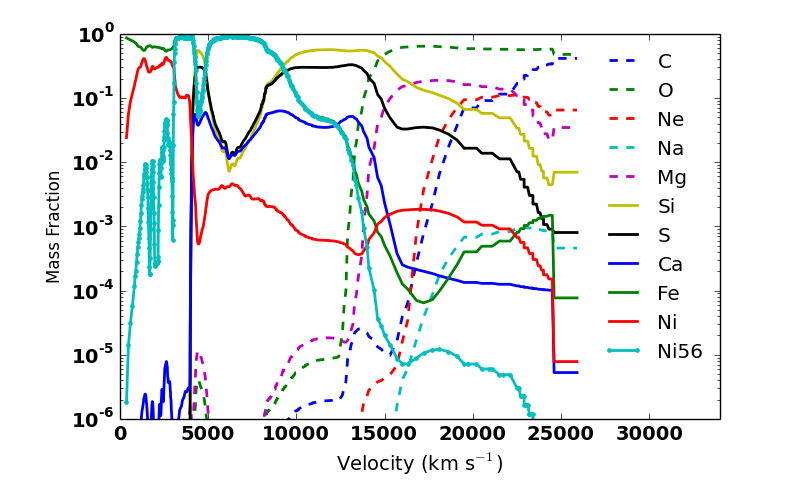}
\caption{The final element distribution of a classical deflagration to
  detonation model.
This is a  delayed-detonation model which 
reproduces the light curves and spectra for Branch-normal supernovae
\citep{hofdd+mol95,HGFS99by02,hoef_review06,gerardy03du04}.
The C/O white dwarf is
from the core of an evolved 5\msol main sequence star.  Through
accretion, this core approaches the Chandrasekhar limit.  An explosion
begins spontaneously when the core has a central density of $2.0
\times 10^9$~\gcm and a mass close to $1.37$~\msol
\citep{h02}. The transition from deflagration  to detonation is
triggered at a density of $2.3 \times 10^{7}$~\gcm. Adapted from \citet{sn11fe_early}.\label{fig:ddt_abunds}}

\end{figure}

In thermonuclear supernovae, SNe~Ia, the explosion energy is provided
by the thermonuclear fusion of the C+O white dwarf to iron group and
intermediate mass elements. The rough structure that any model for a
SN~Ia must reproduce is shown in Figure~\ref{fig:ddt_abunds}. However,
the optical display seen by astronomers is not due to the thermal
energy produced by the thermonuclear fusion of the explosion. While
this energy unbinds the star and produces the kinetic energy of the
explosion, the initial high density and compact radius of a white
dwarf means that it is opaque to radiation until it has expanded in
radius by about a factor of a million. This means the volume has
increased by $10^{18}$ and thus all the stored thermal energy has been
exhausted in $p\,dV$ work. Thus, the optical display for SNe~Ia comes
not from the fusion itself, but rather from the radioactive decay of
\nni, where the $\gamma$-rays and positrons are thermalized and
produce the optical light curves and spectra. 

It is important to understand that the thermonuclear explosion of a
nearly Chandrasekhar white dwarf induces a form a stellar amnesia
\citep{hoef_review06} due to the nuclear physics of the initial progenitor. At
high densities the material will burn to the iron group, producing
\nni, or, if the densities are high enough, electron capture will be
significant and non-radioactive iron group elements will be produced
in the central regions. 

\begin{figure}
\begin{center}
\includegraphics[scale=0.17]{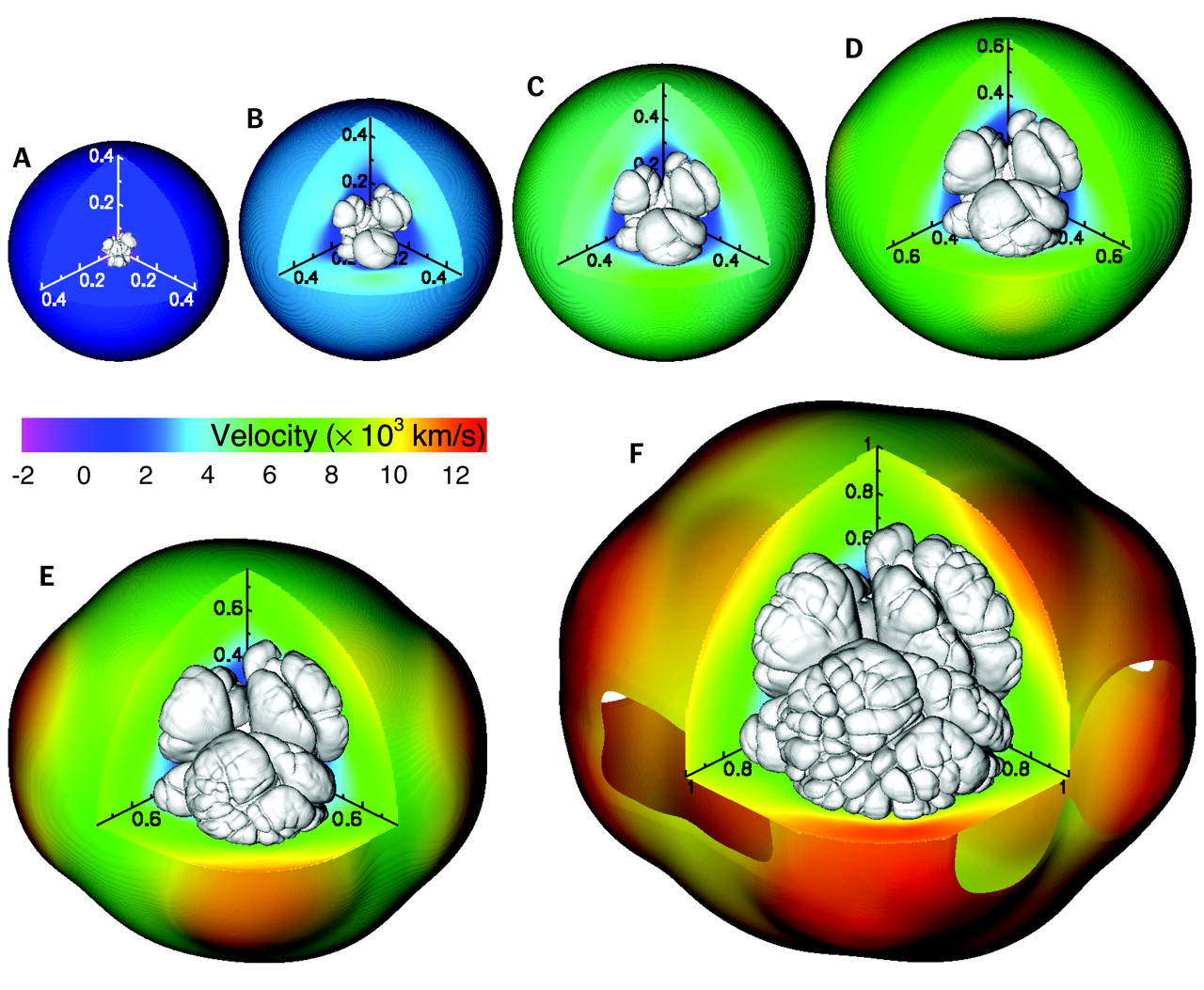}
\end{center}
\caption{3-D models: Pure deflagration leads to  low energy explosion,
  and lots of clumps of unburned 
  material particularly near center. Adapted from
   \citet{Gamezoetal03}.
\label{fig:gamezo_def}
}
\end{figure}

In addition, the explosion itself is complex, in all scenarios it
begins with a subsonic burning phase (deflagration). However, Rayleigh
Taylor instabilities will lead to a well-mixed distribution  of the
elements in contrast to what is observed in SNe~Ia spectra
\citep{Gamezoetal03}. This behavior is shown in
Figure~\ref{fig:gamezo_def}. The favored solution to this problem is
the deflagration to detonation transition (DDT) scenario, where the
explosion begins as a deflagration, allowing the material to pre-expand,
but the deflagration transitions to a detonation at some density
\citep{khok91a,khok91b,khok91c,khok89,K93c,gko04a,gko05}.  The
detonation shock wave travels both forward and backward through the
star burning any mixed unburned material and producing a layered
structure. Figure 4 shows one realization of the DDT model. Several variations on this scenario exist, including the
gravitationally confined detonation \citep{PCL04,meakin09,jordangcd12}
and the pulsating reverse detonation
\citep{bravo_3d_ddt08,bprd07,bravo_prd09b,bravo_prd09a}. While the
deflagration to detonation transition occurs in terrestrial situations
where the burning occurs in a confined region, with walls for the
pressure waves to reflect off of, it is unknown if it naturally occurs
in the unconfined stellar medium, but see Ref.~\cite{PGO11}. 

\begin{figure}
\begin{center}
\includegraphics[scale=0.3]{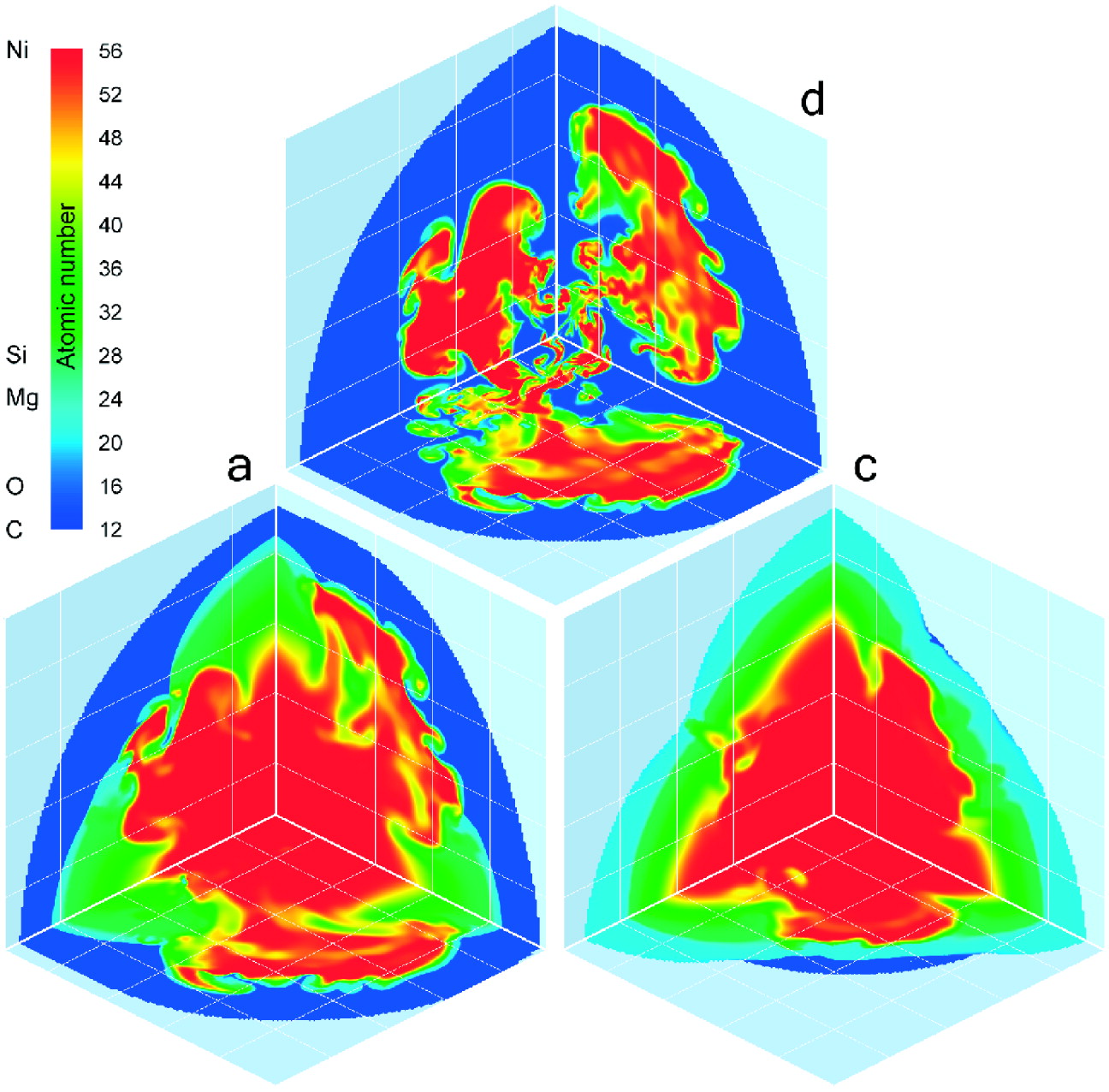}
\end{center}
\caption{3-D models: Delayed detonation, the initial deflagration
  phase  allows the star to 
  pre-expand.  The detonation ``sphericizes'' the incomplete burning
  left from the  
  deflagration. Adapted from  \citet{gko05}.} 
\end{figure}

\begin{figure}
\centering
\includegraphics[scale=.3]{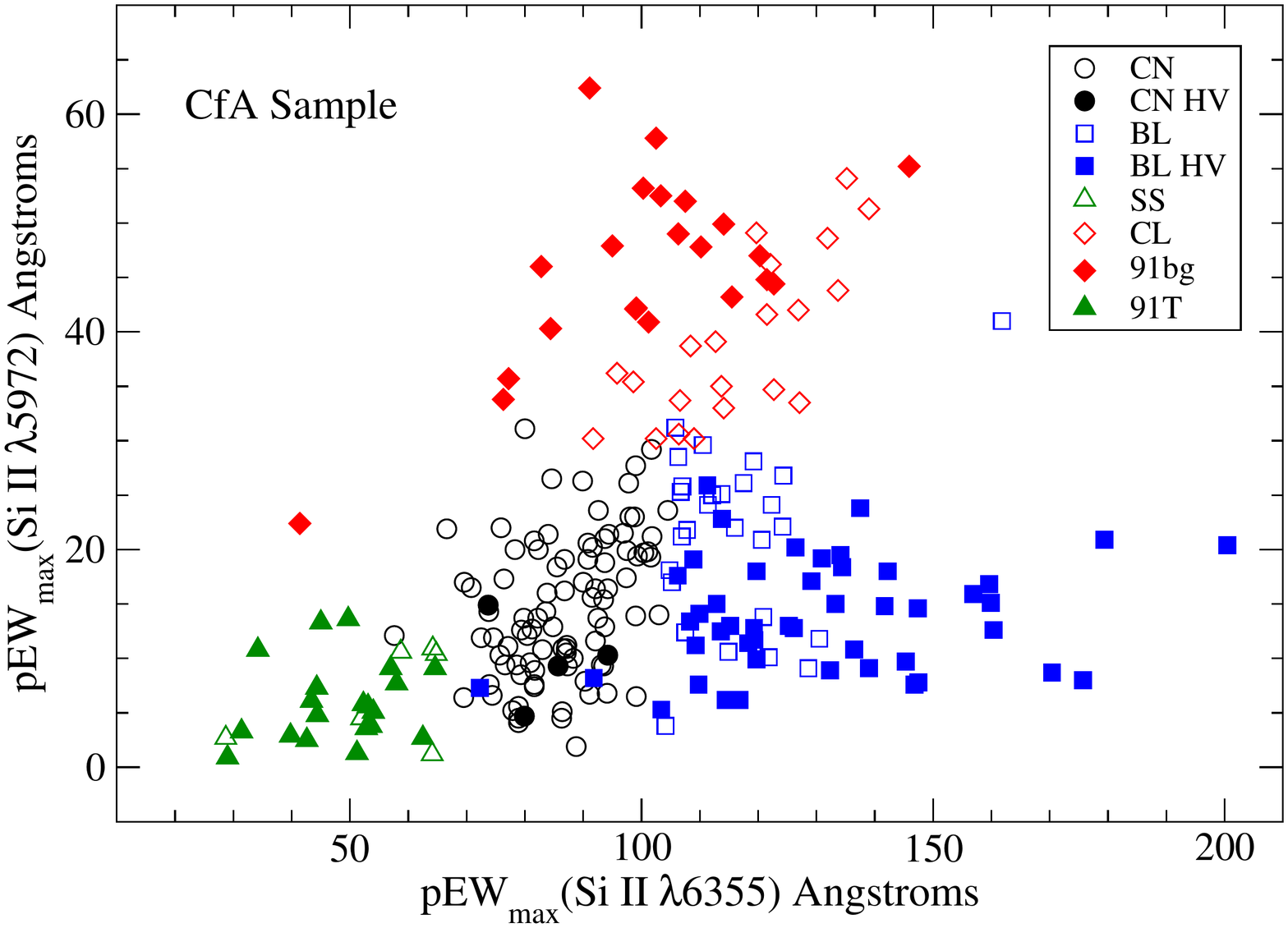}
\caption{Branch et al. diagram of the Si II pseudo-equivalent widths
  based on measurements of CfA spectra published by
   \citet{blondin12}.
  The CN, CL, SS, and BL classes are 
  indicated by the different symbols.  High velocity (HV)
  SNe \citep{xwang09}, which are essentially the same as the 
  HVG class \citep{benetti05} are concentrated mostly among the BL
  objects.  1991bg-like events correspond the CL class, and 1991T-like
  events to the SS class.
\label{fig:wvsw}
} 
\end{figure}

At first glance, SNe~Ia seem remarkably homogeneous in their
observational characteristics.  Nevertheless, observations 
carried out since the 1980's have increasingly revealed
a widespread diversity in spectra and light curves
requiring a whole new understanding of the field. Empirically, considerable
order was
brought to the understanding of SNe~Ia with the
development of the Phillips relation
\citep{philm15,philetal99,goldhetal01}, which is understood as due to a
variation in the total amount of radioactive nickel produced in the
supernova causing higher temperature and hence opacity variations
which leads to variations in the diffusion time.  The correlation in
the brightness (nickel mass) and the diffusion time leads to the
Phillips relation \citep{kmh93,nugseq95,KW07}. Yet, while the light curve
shape relation allows us to use SNe~Ia as standard candles, it does
not explain all of the observed diversity.  

This diversity observed on top of the Phillips relation is sometimes 
generically referred to as the second parameter problem, and
 is partially captured in the work of Branch \etal
\citep{branchcomp105,branchcomp206,branch_pre07,branch_post07,branchcomp509}
who plotted pseudo-equivalent widths of the Si~II $\lambda 6355$
and 5970 features against each other (see Fig~\ref{fig:wvsw}).
Branch \etal used this diagram to group SNe~Ia into four 
classes: core normals (CN), cools (CL), shallow silicon (SS), and
broad line (BL). Using a different approach 
\citet{benetti05} 
arrived at similar classes: Faint (overlapping with CL), High Velocity
Gradient or ``HVG'' (overlapping with BL), and Low Velocity
Gradient or ``LVG'' (overlapping with CN and SS).
Some of this variation has been ascribed to asymmetrical
explosions \citep{maedanature10,maund10a}, and
asymmetric distributions of both
iron group elements (including radioactive nickel) as well as of
intermediate mass elements are possible.  

Since
the total amount of radioactive nickel production generally explains
the Phillips relation, it may not be too far afield to expect that
variations in the zero age main sequence  mass (ZAMS) of the
progenitor, its primordial 
metallicity and the history of the binary system, may well account for
much of the ``second parameter'' diversity described above. 

\begin{figure}
\centering
\includegraphics[width=.85\textwidth,angle=180]{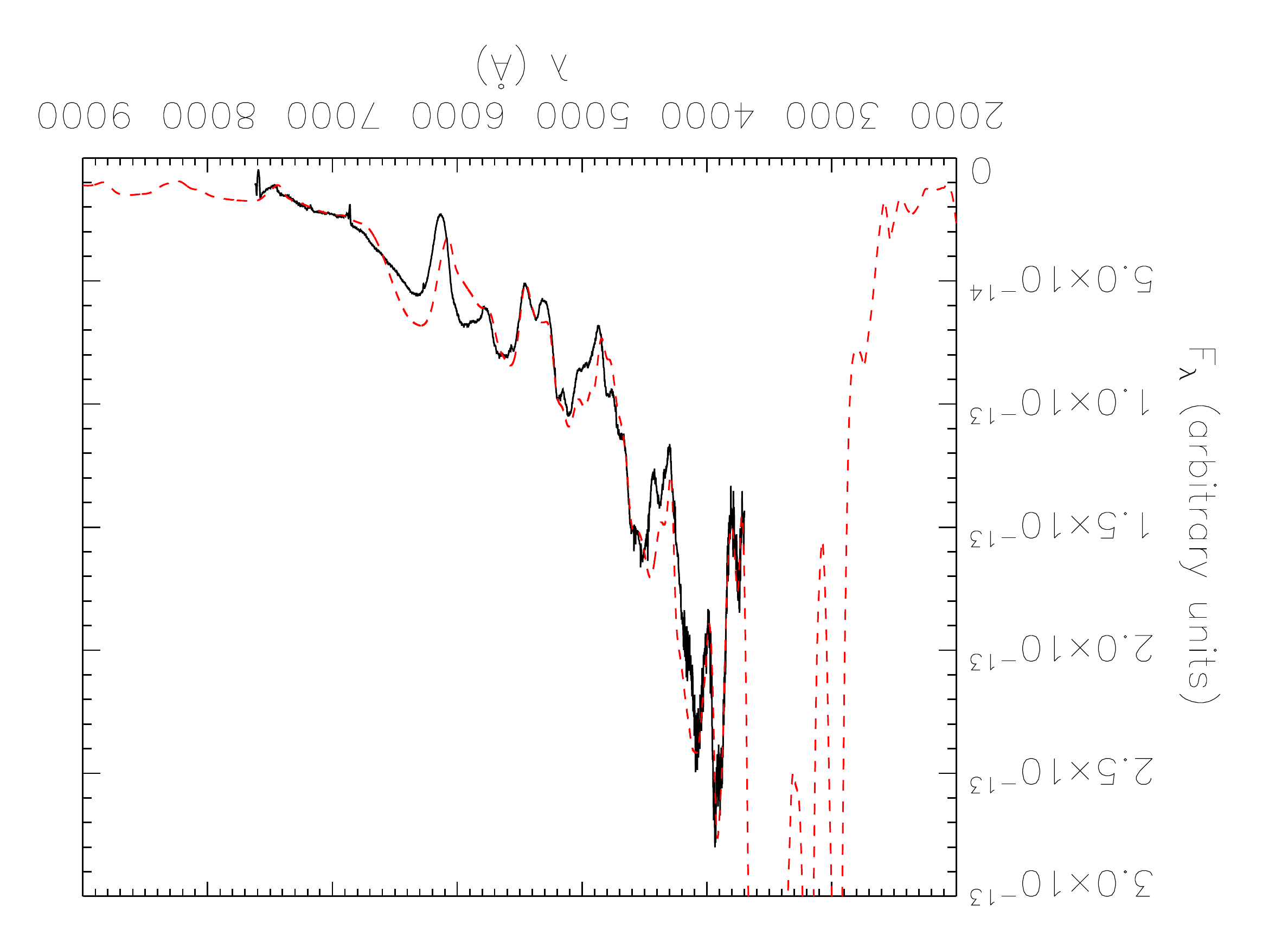}\\[10pt]
\caption{A detailed NLTE calculation of the model W7 compared to the
  observed spectrum of SN 1994D March 21. The observed spectrum has
been corrected for redshift assuming a velocity of 448~\kmps\ and a
reddening of $E(B-V) = 0.06$. Adapted from \citet{bbbh06}.
\label{fig:w7_94D}
}
\end{figure}

\begin{figure}
\begin{center}
\includegraphics[scale=0.45]{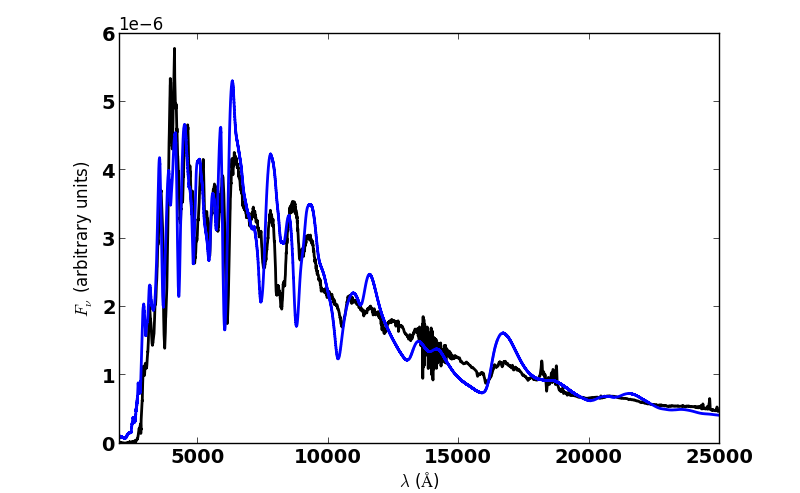}
\end{center}
\caption{Detailed NLTE spectrum of delayed detonation model compared to the
maximum light spectrum of SN 2011fe. The observed spectrum covers the
entire wavelength range from the UV to the IR. Adapted from
\citet{sn11fe_early}. 
\label{fig:sn11fe}
}
\end{figure}

In spite of the detailed diversity of SNe~Ia, they remain important
cosmological probes and their basic layered structure well reproduces
the observed detailed spectra for normal
SNe~Ia. Figure~\ref{fig:w7_94D} shows a detailed non-local
thermodynamic equilibrium (NLTE)  spectral
calculation of the parameterized W7 model compared to the observed
spectrum of the core normal SN~1994D and Figure~\ref{fig:sn11fe} shows
an extremely detailed NLTE
calculation of a standard delayed 
detonation model compared to the full UV--IR spectrum of the nearby,
normal SN~2011fe. 

Observations in the
21st Century have seen the discovery of an uncomfortably 
large number of peculiar ``classes'' of SNe~Ia identified by their prototypes: 
{\it 2000cx} (rare, photometrically-peculiar events that do not
follow the Phillips relation, showing a
rise time typical of a SN Ia, but with an unusually slower decline 
and high photospheric temperature
\citep{li00cx01,candia_optical_2003,Silverman13bh13}); 
{\it 2001ay} (a BL-HVG event with an extremely slow decline
rate but with an apparently modest $^{56}$Ni yield of 0.6 solar
masses \citep{krisc01ay11,b01ay}); 
{\it 2002cx} (events that are spectroscopically similar to 
normal SNe Ia, but have lower maximum-light velocities, 
low luminosities for their decline rates, yet generally hotter photospheres
\citep{li02cx03,Phil05hk07,foley_Iax_13}); 
{\em 2002ic} (SNe~Ia-like 
events with a strong CSM interaction
\citep{ham02ic03,wang02ic04,kotak02ic04,deng02ic04,cch02ic04,han06a,
ben02ic06,dilday11kx12}); and
{\it 2006bt} (SNe Ia with broad light curves like a hot, luminous event
and lacking a prominent secondary 
maximum in the near-IR, but displaying spectra at maximum similar to those of 
low-luminosity SNe Ia \citep{foley06bt10,maguire10ops11}).
Moreover, several SNe~Ia (2003fg, 2006gz, 2007if, 2009dc) 
have been observed whose brightness and light curve
shape have led them to be classified as super-Chandrasekhar explosions
\citep{howell03fg06,jbb07,hicken06gz07,ofek06gy07,scalzo07if10,tanaka09dc10,silverman09dc11}
which may be due to double degenerate explosions where the mass of the
binary exceeds a Chandrasekhar mass, or possibly due to supermassive
white dwarfs due to rotational support \citep{YL05a,YL04c}. 

In fact this wide range of diversity has led to the suggestion that
the parameter responsible for the second parameter variation is the
mass ejected in the explosion itself. This is due either to dynamical
mergers of binary white dwarfs
\citep{pakmornat10,pakmor11,pakmor12a,dong_merge14} or due to pure
deflagration leading to a bound remnant with low ejected mass
\citep{jordan_sublum12,kromer02cx13,fink_def14}.

While these paths may in fact exist in nature, even among the wide
variety of observed supernovae, there is opportunity for the
Chandrasekhar mass scenario to explain some of the observed
diversity. Particularly, pulsational delayed detonations (PDDs) allow
for variation in the \nni \textit{distribution} that explain
deviations from the Phillips relation. 

\begin{figure}
\begin{center}
\leavevmode
\includegraphics[scale=0.5]{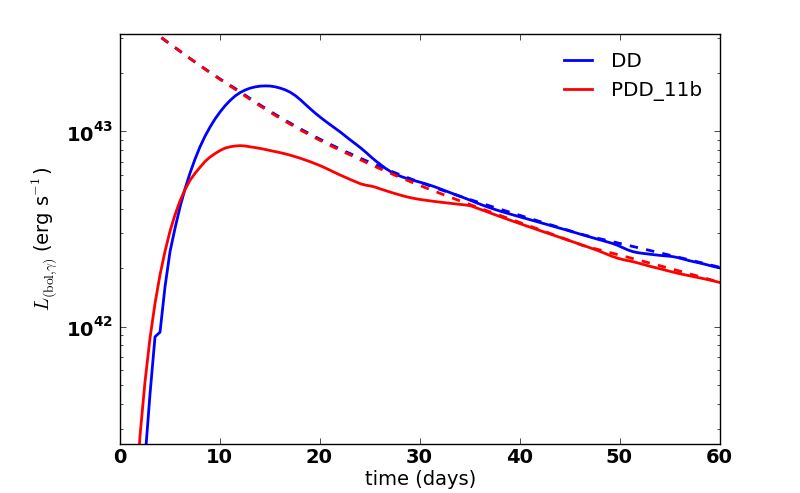}
\caption{The Bolometric light curve of a classic delayed detonation
  model, compared to the pulsational delayed detonation model used for
  SN~2001ay. The dashed  lines show the instantaneous gamma-ray luminosity
  used in Arnett's law. Adapted from \citet{b01ay}.\label{fig:01ay_lc}
}
\end{center}
\end{figure}

\begin{figure}
\begin{center}
\leavevmode
\includegraphics[width=0.90\textwidth,angle=0]{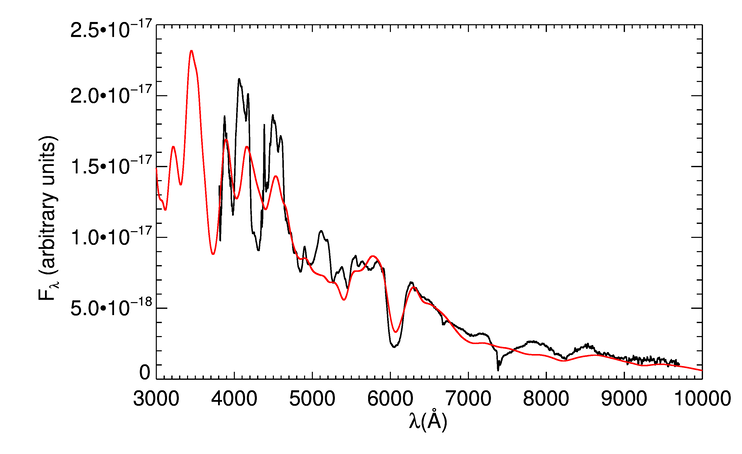}
\end{center}
\caption{Pulsating Delayed Detonation.
SN 2001ay, Max Light, $M_V $ = -19.07 mag. Adapted from \citet{b01ay}.
\label{fig:01ay_spec}
}
\end{figure}

For example, SN~2001ay, the slowest known decliner, was significantly
underbright  for its decline rate \citep{krisc01ay11,b01ay}. By
increasing the C/O ratio, and assuming a PDD we were able to move the
nickel distribution  further out, increasing the kinetic energy and
thus, the amount of $p\,dV$ work done, leading to a slow decline
ratio, normal brightness, and the observed fast spectra \citep{b01ay}. The
bolometric light curve of the model is shown in
Figure~\ref{fig:01ay_lc} and the detailed NLTE synthetic spectrum is
compared to the observations in Figure~\ref{fig:01ay_spec}.

\begin{figure}[h]
\centering
\includegraphics[scale=0.3]{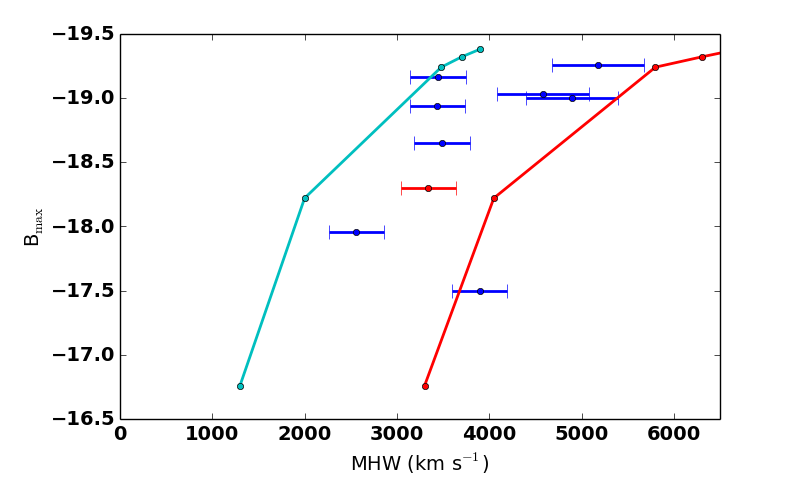}
\caption[]{Mean Half Width, MHW, for the $1.6 \mu$m feature for 
  SNe~Ia (blue) and SN~2012Z (red). 
In addition, the MHW is given for theoretical models of the series 5p0z22 
with (left line) and without mixing (right line) \citep{HGFS99by02}.
Adapted from  \citet{stritz12Z14}.
\label{hwni}
}
\end{figure}

Similarly, the SN~Iax class supernova 2012Z, can be explained by a
PDD in a Chandrasekhar mass white dwarf, where the burning to the iron
group takes place almost exclusively during the deflagration phase,
leading to a central non-radioactive core, some \nni mixing during the
fallback of the bound shell, but the layered structure characteristic
of a detonation in the intermediate mass elements, as well as for the
low velocity spectra with narrow lines, indicating a small
differential spread in velocities
\citep{stritz12Z14}. Figure~\ref{hwni} shows the mean half widths of
the $1.6\mu$m Fe~II feature for a variety of normal and SNe~Iax
supernovae.

Both the models of SN~2001ay and SN~2012Z, show that while the primary
understanding of the Phillips relation is the correlation of the total
mass of \nni produced in a Chandrasekhar mass explosion, additional
variation can be accommodated by variations on the spatial distribution
of \nni, leading to Chandrasekhar mass explosions that do not obey
the Phillips relation.

\begin{figure}
\centering
  \includegraphics[scale=0.5]{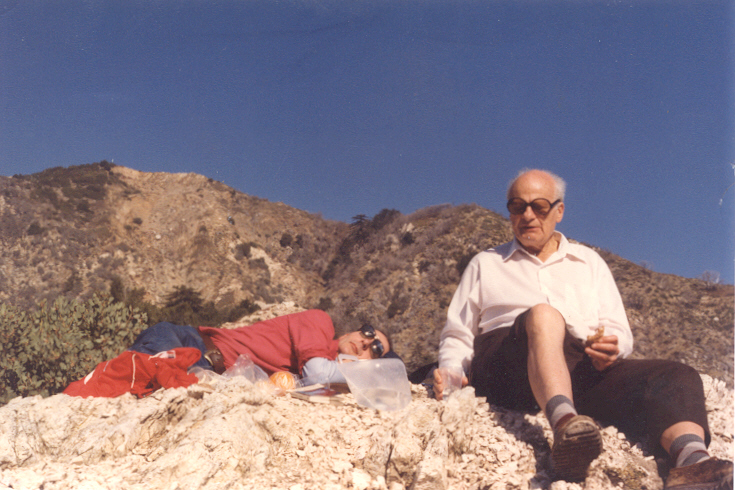}
  \caption{Gerry Brown  and Hans Bethe relaxing in the San Gabriel Mountains
    during a 1982 visit to Caltech. Photo credit: Jerry Cooperstein.
\label{fig:BB82}
}
\end{figure}

There does remain a question of whether there are enough white dwarfs
in binary systems to grow to the Chandrasekhar mass. Calculations of
supernovae rates suggest that including both the single degenerate
channel and the double degenerate channel still produces too few
supernovae compared to the observed galaxy-cluster rate
\citep{claeys14}. 
Chandraskehar mass WD explosions are triggered by compressional heating near
the WD center. Because the compressional heat release increases
rapidly towards the Chandrasekhar mass, exploding stars should have a very
narrow range in masses \citep{hk96}.  The donor star may be either a
red giant or a main sequence star, a
helium star, or the accreted material may originate from a tidally
disrupted WD \citep{WI73,Piersanti03}.  
We differentiate between  dynamic merger models where
a prompt explosion occurs on a dynamic timescale due to heating of
merging material \citep[often called violent mergers][]{kromer_merge13,pakmor12a,pakmor11} and a secular merger in which the matter of the
disrupted companion is accreted by the primary WD on a 
quasi-hydrostatic time-scale. The former leads to an explosion of a
relatively low density configuration. The latter might share many
characteristics with the standard high-density, single-degenerate
$\Mch$ explosion models. 
Efforts have been made to expand the progenitor distribution by
including sub-Chandrasekhar explosions
\citep{ruiter09,ruiter11,ruiter13,ruiter14}. Sub-Chandrasekhar mass
explosions are triggered by helium detonation which produce iron
group elements at the surface making the spectra either too blue
\citep{HKWPSH96,nughydro97} or too red
\citep{kromersub10,pakmornat10,simsub10}. Others have studied channels
where two white dwarfs collide in globular clusters or multiple
systems
\citep{rosswog09,raskin09,hawley12,garciasenz13,kushnir13,dong_merge14}. 
While there is some evidence that the
classical red giant mode of the single degenerate channel may be rare
\citep{weidong_last,chomiuk11fe12,Edwards2012,SP_SNR12} there are
uncertainties on the nature of the environment as well as
uncertainties on the nature of the progenitor white dwarf
\citep{Rosanne_Muk12}. Additionally, there is some solid evidence for
the single degenerate scenario
\citep{dilday11kx12,mccully14a}. 
The study of delay time distributions (DTD) also somewhat favors the double
degenerate scenario \citep{maoz10a,maoz10b,MM12} in that the observed DTD seems
to be proportional to $t^{-1}$. In the single degenerate scenario, the
DTD should decline sharply 
after a few billion years since for longer times the primary will have
smaller main sequence mass and hence produce lower mass white
dwarfs. However the evidence based solely on delay times is not
conclusive \citep{MMN14,greggio10}. 
Thus, while the total mass ejected in the explosion may be a parameter
in some SNe~Ia events, it is interesting to see just how much of the
observered diversity may be explained within the Chandrasekhar mass
scenario  by variations in the \nni distribution.

\section{Conclusions}

While the Phillips relation implies strong homogeneity, which is well
accounted for in the Chandrasekhar mass model combined with
fundamental nuclear physics, it is unclear just how much of the observed diversity can
be accommodated in the Chandrasekhar mass paradigm. Nevertheless, some
peculiar SNe~Ia that don't obey the Phillips relation can in fact be
modeled within the Chandrasekhar mass paradigm and fit many of the
observations. This does not mean that nature does not take advantage
of other channels available.

Finally, Figure~\ref{fig:BB82} shows Gerry and Hans as I fondly remember them.

\section*{Acknowledgments}
I wish to acknowledge my collaborators Peter Hauschildt, Peter
Hoeflich, Mark Phillips,  and David Branch for much advice and helpful
discussions. I thank Mark Phillips and Chris Burns for helping to
construct Figure~\ref{fig:wvsw}. I 
also wish to thank 
my Nuclear Theory Group friends who supported me during nearly a decade at
Stony Brook. This incomplete list includes Jerry Cooperstein, Tom
Ainsworth, Hans Hansson, Karen Kolehmainen, Prakash, Dany Page, and
Jim Lattimer.  The work has been supported in part by support for
programs HST-GO-12298.05-A, and HST-GO-122948.04-A provided by NASA
through a grant from the Space Telescope Science Institute, which is
operated by the Association of Universities for Research in Astronomy,
Incorporated, under NASA contract NAS5-26555.  This work was also
supported in part by the NSF grant AST-0707704 and 
grants SFB 676 and GRK 1354 from the DFG.  This research
used resources of the National Energy Research Scientific Computing
Center (NERSC), which is supported by the Office of Science of the
U.S.  Department of Energy under Contract No.  DE-AC02-05CH11231; and
the H\"ochstleistungs Rechenzentrum Nord (HLRN).  We thank both these
institutions for a generous allocation of computer time.

\section*{References}

\bibliography{apj-jour,mystrings,refs,local,baron,sn1bc,sn1a,sn87a,snii,stars,rte,cosmology,gals,agn,atomdata,crossrefs}

\end{document}